# Optimizing the Ullmann coupling reaction efficiency on an oxide surface by metal atom addition

*Mikel Abadia (1,2,†), Ignacio Piquero-Zulaica (2,3, †), Jens Brede (1), Alberto Verdini (4), Luca Floreano (4), Johannes V. Barth (3), Jorge Lobo-Checa (5,6), Martina Corso (1,2) and Celia Rogero (1,2)*

(1) Centro de Física de Materiales (CSIC-UPV/EHU), Materials Physics Center MPC, Paseo Manuel de Lardizabal 5, E-20018 San Sebastian, Spain

(2) Donostia International Physics Center (DIPC), Paseo Manuel de Lardizabal 4, E-20018 Donostia-San Sebastian, Spain.

(3) Physics Department E20, Technical University of Munich (TUM), 85748 Garching, Germany

(4) Instituto Officina dei Materiali (CNR-IOM), Laboratorio TASC, Trieste, Italy

(5) Instituto de Nanociencia y Materiales de Aragón (INMA), CSIC-Universidad de Zaragoza, Zaragoza, 50009, Spain

(6) Departamento de Física de la Materia Condensada, Universidad de Zaragoza, 50009, Zaragoza, Spain.

(†) M.A. and I.P.-Z. contributed equally to this work

ABSTRACT. The bottom-up synthesis of carbon based nanomaterials directly on semiconductor surfaces allows to decouple their electronic and magnetic properties from the substrates. However, the lack of reactivity on these non-metallic surfaces hinders or reduces significantly the yield of these reactions. Such hurdles practically precludes transferring bottom-up synthesis strategies onto semiconducting and insulating surfaces. Here, we achieve a high polymerization yield of terphenyl molecules on the semiconductor $TiO_2(110)$ surface by incorporating cobalt atoms as catalysts in the Ullmann coupling reaction. Cobalt atoms trigger the debromination of 4,4"-dibromo-*p*-terphenyl (DBTP) molecules on $TiO_2(110)$ and the formation of an intermediate organometallic phase already at room-temperature (RT). As the debromination temperature is drastically reduced, the homo-coupling temperature is also significantly lowered, preventing the desorption of DBTP molecules from the $TiO_2(110)$ surface and leading to a radical improvement on the poly-*para*-phenylene (PPP) polymerization yield. The universality of this mechanism is demonstrated with an iodinated terphenyl derivative (DITP), which shows analogous dehalogenation and polymerization temperatures with a very similar reaction yield. Consequently, we propose to use minute amounts of metal catalyst to drive forward generic bottom-up synthesis strategies on non-metallic surfaces.

**INTRODUCTION**.

Throughout the last decade, on-surface chemistry has proven to be an extraordinary tool for building $sp^2$ bond based carbon nanostructures with unprecedented atomic precision [1–4]. Motivated by the fact that such carbon structures, hard to synthesize by common wet chemistry methods, have prominent electronic properties deserving implementation into electronic devices, the interest of exploring surface induced molecular reactions has grown exponentially[5–9].

The most successful on-surface reaction pathway for controllably growing $sp^2$ carbon nanostructures is the well-known Ullmann coupling reaction due to its relative simplicity, predictability and high reaction efficiency. Therefore, it has been accepted as the most promising reaction pathway for designing carbon based nanostructures for molecular electronics[2,10–12]. However, bringing such materials into nanodevices remains a challenge, as the surface assisted Ullmann reaction is mostly employed in ultra-high vacuum (UHV) conditions on single crystal noble metal substrates (gold[13], silver[14], and copper[15]) given their high catalytic capability. Nevertheless, these on-metal generated nanostructures are impractical, since they remain electronically coupled to the underlying catalyzing substrate. Thus, the decoupling of the synthesized nanostructures must be achieved afterwards for their implementation into devices[16], which is currently performed by inconvenient post growth transfer methods.

The on-surface synthesis of molecular nanostructures directly on semiconductors and insulators would allow to overcome such fundamental problems and to bring such materials into devices. To date, successful reports on this approach are very scarce given the limited catalyzing activity of these substrates[17–20]. Among those, we demonstrated in previous works[21,22], that the on-surface Ullmann coupling reaction can be directly catalyzed by semiconductor surfaces such as $TiO_2(110)$, allowing the polymers frontier bands to sit in between the electronic band gap of the substrate, decoupling their properties. In particular, we showed that the $TiO_2(110)$ surface is capable of catalyzing Ullmann coupling reactions exploiting undercoordinated sub-surface interstitial Ti atoms, which thermally diffuse to the surface and act like a catalyst. Recently, Zuzak *et al.*[23] have followed the same procedure to synthesize different types of nanographenes as well as the well-known 7 armchair graphene nanoribbon (7-AGNR) on a reduced $TiO_2(110)$ sample. Before them, Kolmer *et al.*[24,25] also demonstrated the on-surface synthesis of 7-AGNRs using a *de novo* synthesized fluorinated precursor on the $TiO_2(001)$ surface[26,27].

The extremely low reaction yields of the nanographenes and GNR structures on semiconductor surfaces is mainly related to two fundamental reasons: i) the lack of surface catalytic capability to induce the molecular dehalogenation, the first step within the Ullmann reaction and, ii) the temperatures required to dehalogenate the molecules are very close to their surface desorption temperature or even their decomposition temperature. In the present work, we demonstrate a new pathway to overcome the above mentioned limitations and drawbacks and significantly improve the Ullmann coupling reaction yield of halogenated molecules on the $TiO_2(110)$ surface. By adding minute amounts of cobalt atoms in the initial stages of the reaction, a dramatic improvement of the yield of polymerization of 4,4"-dibromo-*p*-terphenyl (DBTP) molecules into poly-para-phenylene (PPP) polymers by almost 300% is observed. The generality of the mechanism is demonstrated by obtaining a similar efficiency for the iodinated DITP molecules. We unambiguously show this by means of synchrotron-based X-ray photoemission spectroscopy (XPS), lab-based angle-resolved photoemission spectroscopy (ARPES), low energy electron diffraction (LEED) and room temperature scanning tunneling microscopy (RT-STM) techniques.

**RESULTS.**

The measurements are performed as follows: on the one hand, a sample consisting of a single layer DBTP deposited on the $TiO_2(110)$ surface is used for control and comparison purposes (from here on, the "control sample"). The molecular adsorption is self limited to a single layer, ensuring the repeatability of the sample preparations[15]. On the other hand, a second sample is similarly prepared, with a single layer DBTP deposited on the $TiO_2(110)$ surface, but here, in a second step, cobalt atoms are thermally evaporated onto the sample kept at room temperature (RT) (from here on, the "cobalt sample"). The temperature evolution of the surface reactants is then systematically measured for both samples, in equivalent conditions, and the differences within the Ullmann coupling reaction pathway are determined by photoemission techniques.

- **Core level line shapes at room temperature**

In order to limit the catalytic contribution of sub-surface Ti interstitial atoms, a nearly stoichiometric TiO$_2$(110) surface is used throughout this work, whereas purposely reduced samples were used in previous studies[21,23]. In fact, the formation of Ti interstitials is associated with sample reduction, whose charge excess is redistributed among a few lattice sites[28,29] and is manifested by the appearance of a new state in the gap (the Defect State, DS, at ~0.9 eV, corresponding to the partial filling of the Ti3$d$ band) and the appearance of a shoulder at the low binding energy, BE, side of the Ti2$p$ core level (corresponding to a nominal Ti$^{3+}$ oxidation state). Thus, we may assume these two spectroscopical features as a probe of the concentration of subsurface Ti interstitials [see Figure S1 in the Supplementary Information (SI)].

The molecular structure of DBTP and the atomic structure of the TiO$_2$(110) are shown in **Fig. 1a** and **b**, respectively. In this same figure, in panels **c** and **d**, we show the XPS measured on Br3$d$ and C1$s$ CLs at RT for the control (in black) and cobalt (in red) samples. The C1$s$ CL peak on the control sample, black spectrum in **Figure 1d**, is deconvoluted with three components, in agreement with previous work[21,22] i.e., two halogen bonded carbons at 285.8eV (C-Br), four carbons in the *para* position of the benzene rings at 284.9eV (C-C), and the remaining twelve carbons of the molecular backbone with the lowest binding energy (BE) at 284.7eV (C-H). The BE of the spin orbit coupled Br$_{3d}$ double peaks shown in black in **Figure 1c** indicate that DBTP molecules remain halogenated after deposition on the TiO$_2$(110) surface[22] (see **Fig. 1b**). The absence of cobalt atoms in the control sample is evident from the absence of the Co$_{2p}$ CL peak in **Figure 1e** (top spectrum).

We find significant changes in the cobalt sample for both Br3$d$ and C1$s$ peaks (red spectra in **Figure 1c** and **d** respectively). The most striking one is the presence of an additional component of the Br3$d$ CL, shifted by 1.3 eV to lower BE (from 70.6 eV to 69.3 eV). Similarly to noble metal surfaces[30], the observed shift is ascribed to the debromination of the DBTP molecules, where bromine atoms remain absorbed on the surface after being detached from their molecular counterpart. In contrast, the debromination of DBPT molecules does not occur before 450K in the control sample, as previously reported[21,22] and evidenced in **Figure 2d**.

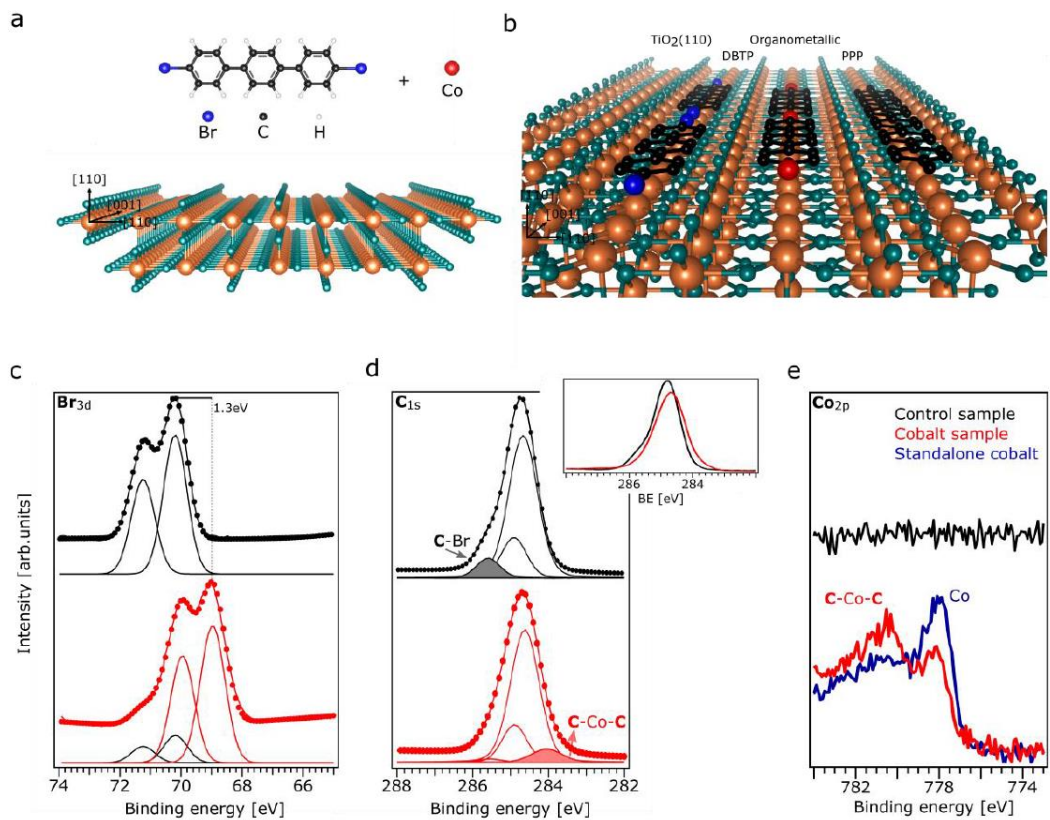

**Figure 1.** The Ullmann coupling reaction constituents are shown in a), i.e., DBTP as molecular precursor, cobalt as external catalyst and the template substrate TiO$_2$(110). b) The DBTP adsorption on the TiO$_2$(110) trenches is shown, as well as the intermediate organometallic phase and the final PPP reaction product. XPS measurements for DBTP on the control (black spectra) and cobalt sample (red spectra) are shown in c), d) and e) for Br3$d$, C1$s$ and Co2$p$ CLs respectively. In the last panel, the Co2$p$ CL spectrum of cobalt atoms as deposited on the clean TiO$_2$(110) surface is included for direct comparison (blue spectrum). The inset in (d) compares the C1$s$ peak shape of control and cobalt samples. Measurements are performed with the samples at RT (see Methods section).

Subtle but consistent differences also emerge when comparing the C1$s$ spectra of both samples. The inset top image of **Figure 1d** shows a decrease in the signal intensity at the high BE side of the red spectra, accompanied by an intensity increase in the low BE side. Interestingly, the overall C1$s$ CL integrated intensity area remains barely unchanged between both spectra, indicating that no molecular desorption occurs upon the addition of cobalt atoms to the sample at RT. The deconvolution of the peak, red spectrum in **Figure 1d**, further accentuates the observed differences. On the one hand, compared to the control sample, the high BE peak at 285.5eV (C-Br) practically disappears, which has recurrently been assigned to the natural evolution of the C1$s$ CL upon the debromination of DBTP molecules[5,14,15,21,22] and is in line with the results derived from the Br3*d* CL signal. On the other hand, the emergence of a new peak component at 284eV points out to the formation of an organometallic intermediate phase (C-Co-C), where cobalt atoms coordinate with the debrominated radical carbon atoms of adjacent molecules to stabilize their charge, as illustrated in **Figure 1b**. We note that, while similar organometallic phases have frequently been reported on metallic Cu or Ag surfaces[15], this is, to the best of our knowledge, the first observation of such a phase on a semiconducting surface. We believe that the excellent registry of the DBTP molecules with the natural trenches of the $TiO_2$(110) substrate[21,24] likely promotes the formation of the organometallic phase via geometric constraint.

The organometallic phase formation can also be seen on the Co2*p* CL spectra shown in **Figure 1e**. The blue spectrum represents the adsorption of cobalt atoms on the clean $TiO_2$(110) surface, with a pronounced peak at 778eV corresponding to its metallic nature. However, in the presence of the DBTP monolayer, the peak shifts towards higher BE due to the charge redistribution in the newly formed organometallic phase (red spectrum). Moreover, the higher oxidation state of the cobalt peak indicates that, apart from stabilizing the radical carbon atoms on the organometallic phase, cobalt atoms also interact severely with the detached bromine atoms. Overall the results evidence the strong chemical interaction of the precursor molecules with cobalt atoms.

- **Temperature dependence of the species**

We further follow the polymerization reaction on both, the reference as well as the cobalt sample by temperature dependent XPS (TD-XPS) measurements that are shown in **Figure 2**. Here, both samples are annealed from 300K (RT) to 620K, while the C1*s* and Br3*d* CLs of DBTP are monitored. Unfortunately, the minute amounts of Co prevents us from visualizing such evolution within a similar timeframe, therefore we select three temperatures to acquire Co2*p* spectra with the necessary statistics. In this way, three mayor spectroscopic fingerprints are singled out after comparing both samples: the molecular debromination temperature difference (already evident at RT for the cobalt sample), the identification of the homo-coupling temperature of the constituent molecules, and the final C1*s* CL signal intensity, directly correlated with the reaction efficiency.

The temperature evolution of the C1*s* and Br3*d* CL on the control sample, **Figure 2a** and **d** respectively, have been extensively analyzed elsewhere[21,22] and the results are summarized as follows: At 475K, the Br3*d* CL signal intensity dramatically drops due to the DBTP debromination and the subsequent desorption of the bromine atoms from the surface. At this temperature, the intensity of the C1*s* CL drops to roughly a quarter of its initial value (see the dashed black line in the intensity profile shown in **Figure 2c**). This demonstrates that a significant amount of DBTP molecules desorb from the surface before polymerizing. The remaining C1*s* signal intensity arises from the successfully polymerized molecules. The distinct highly dispersive band resulting from the newly formed PPP polymer is measured with ARPES and shown in **Figure 3a**.

The observed low reaction efficiency on the control sample (≈25%) responds to two main reasons[22]: one is that intrinsically, there are very few sites that can catalytically activate the molecular debromination on the barely reduced $TiO_2(110)$ surface. These are mostly surface defects with excess of accessible charge, often produced by standard UHV sample preparation procedures. In the present case the presence of such defects is deliberately minimized to preserve the pristine properties of $TiO_2(110)$ surface (non reduced), so that only few interstitial Ti atoms, capable of diffusing to the sample surface, can induce the debromination of the DBTP molecules. The second reason is that the thermally activated outdiffusion of Ti interstitials[31] takes place in the same temperature range of DBTP debromination and desorption temperature. Simply put, many DBTP molecules are re-evaporated from the surface before the reaction can take place.

In the following we show how the lowering of the debromination temperature (by 150K) brings exceptional consequences into the Ullmann coupling reaction yield on the $TiO_2(110)$ surface. The TD-XPS of the cobalt sample presents clear differences in both core level lines. The decrease in intensity of the C1*s* CL signal, **Figure 2b**, occurs much earlier at 420K, i.e., 55K lower than on the control sample. Notably, the total signal intensity in **Figure 2c**, is almost three times higher compared to the control sample (about 60% of the RT signal), even at more elevated temperatures. This proves that the molecular desorption is considerably reduced. Since the cobalt atoms activate the debromination of the DBTP molecules already at RT, most of the intermolecular homo-coupling can then occur earlier (by 55K), well below their surface desorption temperature.

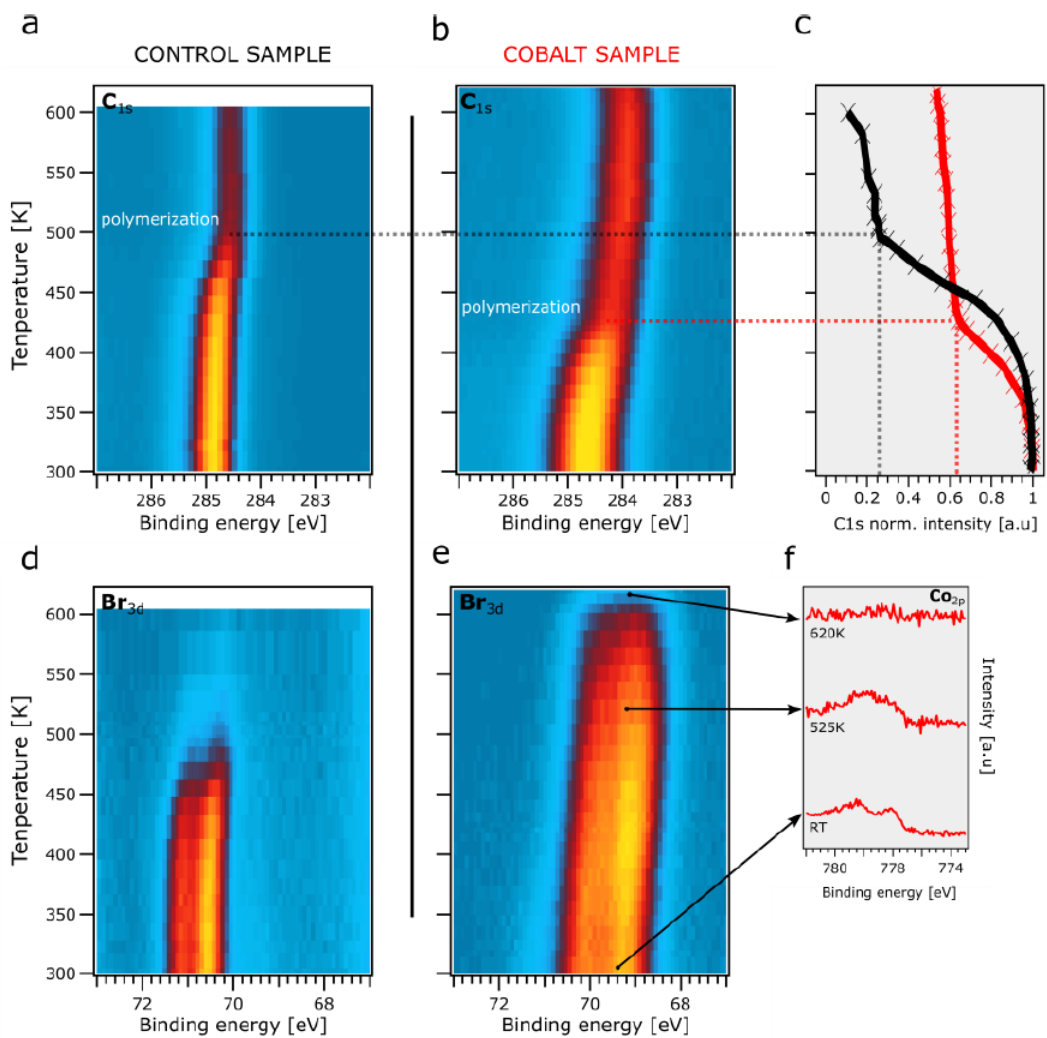

**Figure 2.** Comparison TD-XPS measurements performed on the C1$s$ and Br3$d$ CLs of DBTP in the control (a and d) and cobalt samples (b and e). Fast XPS acquisition was performed while heating the sample from 300K to 620K with a constant linear ramp of 7.5K/min. In c the normalized integrated area of C1$s$ CL is plotted against temperature for both samples. Finally, f shows the Co2$p$ CL peak of the cobalt sample at three selected temperatures: RT, 525K and 620K.

The temperature evolution of the Br3$d$ CL is also considerably different on the cobalt sample. The surface desorption of bromine atoms happens at 550K, i.e., 75K higher than on the control sample (475K). We correlate this higher desorption temperature of the bromine atoms to the presence of cobalt atoms on the surface leading to a Br-Co complex, as evidenced by the change in Co2$p$ lineshape in **Figure 2f**. The high resolution XPS spectra of Co2$p$ CL shows the presence of cobalt atoms on TiO$_2$(110) at RT and 525K but not at 620K, in agreement with the changes observed in the C1$s$ and Br3$d$ CLs. We interpret the temperature evolution of the Br3$d$ and Co2$p$

CLs as the formation of a Br-Co complex that desorbs at around 600K. Note that the C1*s* intensity remains constant at this temperature, proving that the homo-coupling reaction has taken place and the polymers remain on the TiO$_2$(110) surface (see **Fig. 1b**).

- **Band structure of the polymers**

So far, core-level spectroscopic characterization gives strong support of the cobalt induced molecular dehalogenation at RT and the homo-coupling of the dehalogenated molecules at 420K. The formation of PPP chains is further confirmed by LEED images obtained at several steps of the reaction, as well as by RT-STM imaging after the 450 K annealing step (see Figure S2 and S3 in SI). Ultimate proof of the successful reaction is the delocalization of the atomic carbon $p_z$-orbitals along the entire polymer. Experimentally, such delocalization manifests in a highly dispersive π-band[15,32] which can directly be measured by ARPES.

In **Figure 3** ARPES intensity maps of the control and cobalt samples are presented. The control sample spectrum is measured after annealing to 475K and the cobalt sample after annealing to 425K. The ARPES mappings in **Figure 3a** corresponds to the PPP band dispersion parallel to the chains, i.e., along the sample [001] direction (see **Figure 3d**). In both cases, the photoemission intensity presents a highly dispersive band with its apex located at 1.45 Å$^{-1}$, in line with the polymer inter-phenyl distance periodicity[21,32].

The improvement of the reaction yield (60% vs 25%) is evidenced by the stronger intensity of the cobalt sample compared to the control sample, highlighted in the energy distribution curves (EDCs) extracted at 1.45 Å$^{-1}$ and shown in **Figure 3b**. Due to the absence of the TiO$_2$(110) surface DS, otherwise located close to -1eV[21,33], the first spectral intensity that emerges in both samples is the VB onset located at ≈-2eV. In the control sample (black spectrum) the feature is rather weak when compared with the pronounced peak present in the cobalt sample (red spectrum). As already indicated, this intensity difference can be directly ascribed to the signals arising from the amount of PPP chains on the surface. Thus, the combination of ARPES and XPS unambiguously shows the benefit of introducing minute amounts of Co for promoting the Ullmann coupling reaction on the TiO$_2$(110) surface.

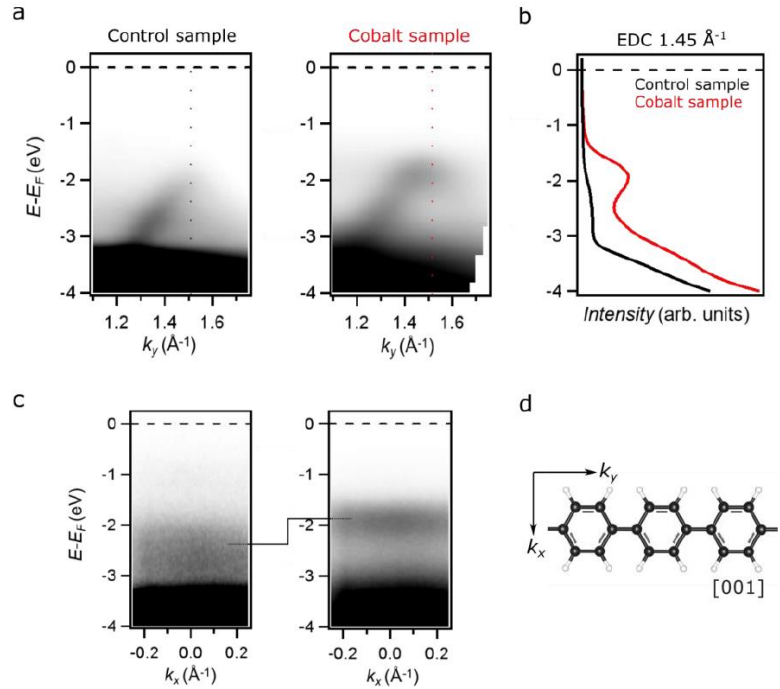

**Figure 3.** Band structure measured by ARPES of PPP polymers grown on $TiO_2(110)$ for both control and cobalt samples after annealing to 475K and 425K, respectively. (a) $E$ vs $k_y$ photoemission intensity maps where the down-dispersive parabolic band along the PPP polymer axis is detected. (b) Comparison of EDC profiles extracted at $k_y$ = 1.45 Å$^{-1}$ for both samples measured in exactly the same experimental conditions. (c) $E$ vs $k_x$ photoemission intensity mapping extracted at $k_y$ = 1.45 Å$^{-1}$) for both samples. (d) Molecular model of PPP along with the wave vector directions of the photoemission measurements.

The presence of Co and Br atoms on the surface induce electronic doping on the polymer chains. In **Figure 3c**, ARPES intensity maps of the band structure perpendicular to the top of the PPP band at $k_y$ = 1.45 Å$^{-1}$ are shown (see **Fig. 3d**). The black line highlights the shift of about 0.2eV of the top of the molecular π-band toward the Fermi level of the cobalt sample with respect to the control one. Piquero-Zulaica *et al.*[34] found such a shift by Scanning Tunneling Spectroscopy (STS) measurements, when zigzag shaped poly-*meta*-phenylene polymers were in contact with Br atoms in metallic surfaces and the effect has been theoretically elucidated by Maier *et al.*[35]. Thus, considering that in the cobalt sample the Br atoms remain on the $TiO_2(110)$ surface up until 600K, we assign the observed band energy shift to the presence of bromine and cobalt atoms surrounding the PPP polymer (see Figure S3 in SI). Note that for the control sample, the bromine atoms have almost entirely desorbed at 475K (**Figure 2d**).

- **Validation of the results with an iodine derivate**

Finally we generalize the use of cobalt atoms as an Ullmann coupling catalyzer on the TiO$_2$(110) surface by expanding this study to the iodine terminated triphenyl sister precursor (DITP). An analogous experimental design to that described so far for DBTP, where a control (DITP) and a cobalt sample (Cobalt and DITP) are prepared and compared. Then, similarly to what it was shown in **Figure 2**, the TD-XPS of the Ullmann coupling and polymerization process of DIPT upon the presence of Co catalysts on the surface is measured. The results are summarized in **Figure 4**.

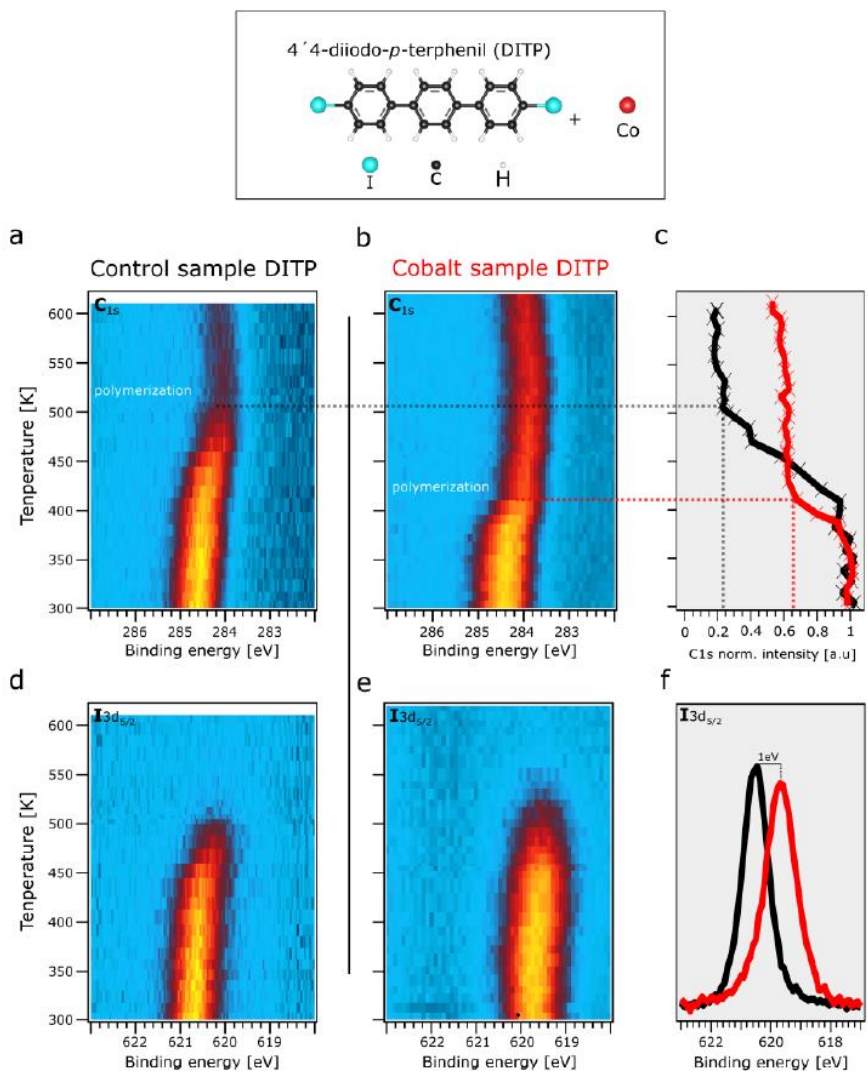

**Figure 4.** Comparison of TD-XPS measurements performed on the C1*s* and I3*d* CLs of DITP in the control (a and d) and cobalt samples (b and e). The temperature is increased from 300K (RT) to 620K. In c the normalized integrated intensity area of C1*s* CL is plotted against temperature for both samples. RT high resolution spectra of I3*d* CL measured in both samples (control sample in black and cobalt sample in red) is shown in f.

The dehalogenation of DITP occurs at RT upon the addition of cobalt atoms, as seen in the high resolution XPS measurements in **Figure 4f**. The 1.0eV shift of the I$3d$ CL from the control (black line) to the cobalt (red line) samples indicates the formation of an organometallic (C-Co-C) phase, demonstrating that cobalt atoms effectively catalyze the dehalogenation of the molecular precursors. This expands the potential use of cobalt as a general approach for implementing the Ullmann coupling reaction on $TiO_2(110)$.

The TD-XPS analysis of the DITP system presents some differences compared to the DBTP system, providing insights into the catalytic behavior of cobalt atoms. The surface desorption temperature of iodine in the presence of cobalt is significantly lower than that of bromine. As shown in **Figure 4e**, iodine atoms desorb from the surface near 500K, while bromine atoms resided on the surface up to 600K in **Figure 2e**. This temperature difference is likely due to the different chemical stability of the cobalt-bromine and cobalt-iodine compounds (see figure S4). Bromine has a higher electronegativity and forms stronger covalent bonds than iodine with the substrate. Thus, the observed temperature differences are assigned to the chemical differences of the cobalt-halide phases formed upon the addition of cobalt atoms[36].

The dehalogenation of DITP on the control sample takes place at 475K, **Figure 4d**, similar to what was observed for DBTP in **Figure 2d**. In contrast, on metallic surfaces like gold, the dehalogenation of iodine occurs at lower temperatures than bromine[37]. This indicates that, on $TiO_2(110)$, without cobalt atoms, the dehalogenation process is primarily determined by the thermal stability of the carbon-halogen bond, rather than by the surface's catalytic capability.

Finally, the polymerization temperature of DITP is identified from the C1s CL core level shift. In the control sample (**Figure 4a**), this shift is observed at 475K and is accompanied by a significant drop in signal intensity to 25% of the initial value, similar to the results obtained for DBTP. In the presence of cobalt, the polymerization temperature of DITP drops to 420K (**Figure 4b**) and the final signal intensity is 60% of the initial value. This, again, is consistent with the results obtained for DBTP in **Figure 2b** and demonstrates that regardless of the halogen atom present in the precursor molecules the Ullmann coupling reaction is dramatically improved by using cobalt atoms as catalyst on the TiO2(110) surface.

**CONCLUSION.**

The use of cobalt atoms is an effective strategy to catalyze the Ullmann coupling reaction of terphenyl derivatives on the $TiO_2(110)$ surface. The yield of their polymerization reaction is practically tripled, and the polymerization temperature is significantly lowered by 55K. The improvement in the Ullmann coupling reaction originates from the strong cobalt-molecule interaction, that promotes the dehalogenation and the formation of an organometallic phase even at RT. This allows to improve the probability of the molecules to polymerize afterwards, which contrasts with the previously studied cases in the absence of cobalt where the molecular dehalogenation temperature practically coincides with their surface desorption temperature. Moreover, we confirm that the presence of bromine and cobalt atoms in the vicinity of the molecules shifts the highly dispersive valence band of the PPP polymer by 200meV towards the Fermi level. Such electronic doping can be used as a simple way to modulate the band energy of conjugated π-bands on the $TiO_2(110)$ surface. In a nutshell, this study presents a new strategy for implementing the on-surface Ullmann coupling reaction with high efficiency in poorly reactive semiconducting or insulating surfaces such as $TiO_2(110)$, opening a promising avenue for synthesizing graphene-based nanostructures such as graphene nanoribbons and nanoporous graphene structures directly on technologically more relevant surfaces.

**METHODS.**

Experiments were carried out in UHV systems at a base pressure of 10–10 mbar. We performed X-ray photoemission spectroscopy (XPS) measurements at the ALOISA beamline of the Elettra Synchrotron (Trieste, Italy). Measurements were performed in transverse magnetic polarization (i.e., close to p-polarization) and normal emission geometry, with the sample at a grazing angle of 4°. The photoemission spectra of Br 3d and C 1s (measured using hν = 500 eV, DE~160meV) and of the valence band (at hν = 140 eV, DE~115meV) were calibrated to the BE of Ti 3p at 37.6 ± 0.05 eV. The Ti 2p3/2 core level was measured using hν = 650 eV, DE~260meV. High-resolution spectra were recorded at RT. The temperature-dependent fast-XPS scans of the Br 3d doublet were measured using a temperature ramp speed of 7.5 K/min.

For preparing a nearly stoichiometric TiO2(110) crystal only two or three sputtering–annealing (Ar+, 1 keV-900K) cycles were performed to obtain a slightly conductive surface but retaining the optical transparency of the crystal. 4,4-Dibromo-p-terphenyl (DBTP) molecules (Sigma-Aldrich, purity higher than ≥88%) were sublimated from a degassed Knudsen cell (molecules degassed several hours/days by heating the source to temperatures slightly below 350 K under UHV conditions) heated to about 375 K to obtain a rate close to 0.33 ML/min. Co was deposited from a rod with commercial e-cell (Focus) at pressures below $6 \times 10^{-9}$ mbar onto the TiO2(110) crystal held at RT.

Scanning tunneling microscopy was carried out at RT using an Omicron VT-STM. Image processing was done with the WSxM software[38].

Angle-resolved photoemission measurements were performed using a Phoibos 150 SPECS high-resolution hemispherical electron analyzer while the sample was cooled to 150 K. He-I$\alpha$ (h$\nu$ = 21.2 eV) radiation was provided by a high-intensity UVS-300 SPECS discharge lamp coupled to a TMM-302 SPECS monochromator. All binding energies given here were referenced to the Fermi level (EF), i.e., BE = 0 = EF.

The model of the atomic structure of the $TiO_2$(110) surface and the chemical structures of DBTP and DITP are done with the VESTA software[39].

**AUTHOR INFORMATION**

Corresponding authors

Correspondence and requests for materials should be addressed to Mikel Abadia and Celia Rogero. (mikel.abadia@dipc.org and celia.rogero@csic.es ).

**ACKNOWLEDGMENT**

M.A. acknowledges support from Prof. Iñigo Atorrasagasti. We acknowledge support from the Basque Departamento de Educación, UPV/EHU (IT-1591-22). This work was funded by the Spanish MCIN/AEI/ 10.13039/501100011033 (PID2019-107338RB-C63, PID2019-107338RB-C64, PID2019-109555GB-I00, PID2020-114252GB-I00), the IKUR strategy, under the collaboration agreement between Ikerbasque Foundation and MPC and DIPC, on behalf of the Department of Education of the Basque Country.

# Supplementary Information:

# Improving the Ullmann coupling reaction efficiency on an oxide surface by metal atom addition

Mikel Abadia[1,2,†], Ignacio Piquero-Zulaica[2,3,†], Jens Brede[,1], Alberto Verdini[4], Luca Floreano[4], Johannes V. Barth[3], Jorge Lobo-Checa[5,6], Martina Corso[1,2] and Celia Rogero[1,2]

[1] Centro de Física de Materiales (CSIC-UPV/EHU), Materials Physics Center MPC, Paseo Manuel de Lardizabal 5, E-200018 San Sebastián, Spain

[2] Donostia International Physics Center (DIPC), Paseo Manuel de Lardizabal 4, E-20018 Donostia-San Sebastian, Spain

[3] Physics Department E20, Technical University of Munich (TUM), 85748 Garching, Germany

[4] Instituto Officina dei Materiali (CNR-IOM), Laboratorio TASC, Trieste, Italy

† M.A. and I.P.-Z. contributed equally to this work.

[5] Instituto de Nanociencia y Materiales de Aragón (INMA), CSIC-Universidad de Zaragoza, Zaragoza, 50009, Spain

[6] Departamento de Física de la Materia Condensada, Universidad de Zaragoza, 50009, Zaragoza, Spain

**Checking the absence of defects on the TiO$_2$(110) surface**

Ti interstitials atoms are naturally formed in TiO$_2$(110) sample as a byproduct of oxygen reduction which is stimulated by the standard cleaning and ordering protocols in UHV (ion bombardment and high temperature annealing), as well as by X-ray irradiation. The excess of charge associated with oxygen vacancies (even the buried ones) is redistributed among a few characteristic subsurface lattice Ti atoms[1]. Such atoms present clear spectroscopic fingerprints either in the Ti2$p$ CL peak, as a shoulder in the lower binding energy site or in the VB, in the form of a pronounced peak around 0.9eV, normally called the defect state (DS) peak. Both spectroscopic features are representative of the defectiveness of the sample, i.e. of its stoichiometry, as highlighted in the comparative spectra shown in Figure S1.

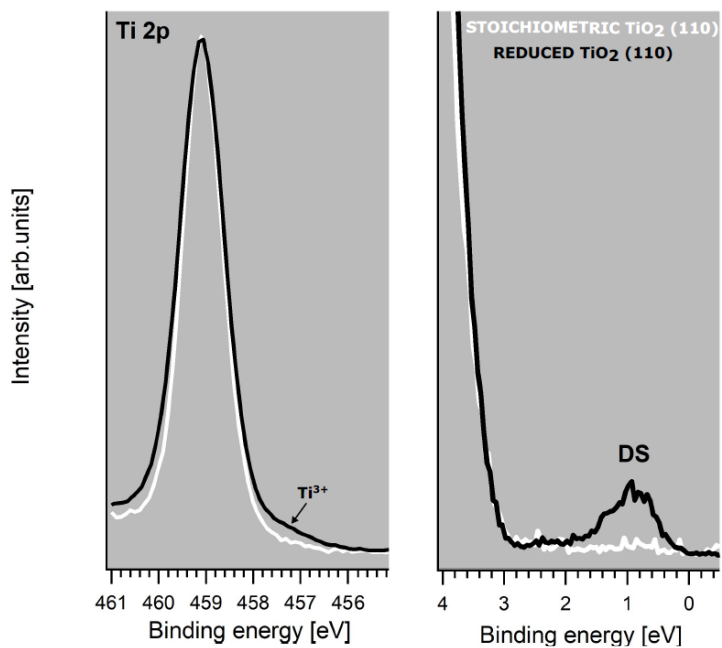

**Figure S 1:** Spectroscopic fingerprints of the stoichiometric (in white) and a reduced $TiO_2$ (110) (in black) surfaces. Ti2$p$ CL (left) and the VB region (right) are shown. In this work we exclusively used samples presenting the white lineshape.

**LEED evolution of the DBTP structure upon Co deposition and subsequent annealing steps**

Figure S2 shows the LEED evolution during the Ullmann coupling reaction of the cobalt sample (panels c-e). Panels a, b serve as a comparison where the pristine clean $TiO_2$(110) and a full monolayer of DBTP on $TiO_2$(110) are shown respectively. For the cobalt sample (panels c-e) a clear evolution of the LEED pattern becomes evident: at room temperature, the organometallic (C-Co-C) chains appear as vertical diffraction lines (panel c, highlighted with green arrows), after annealing to 420K, the characteristic diffraction lines of PPP polymers show up (panel d, highlighted with red arrows) and such PPP lines persist even after 600K annealing step (panel e).

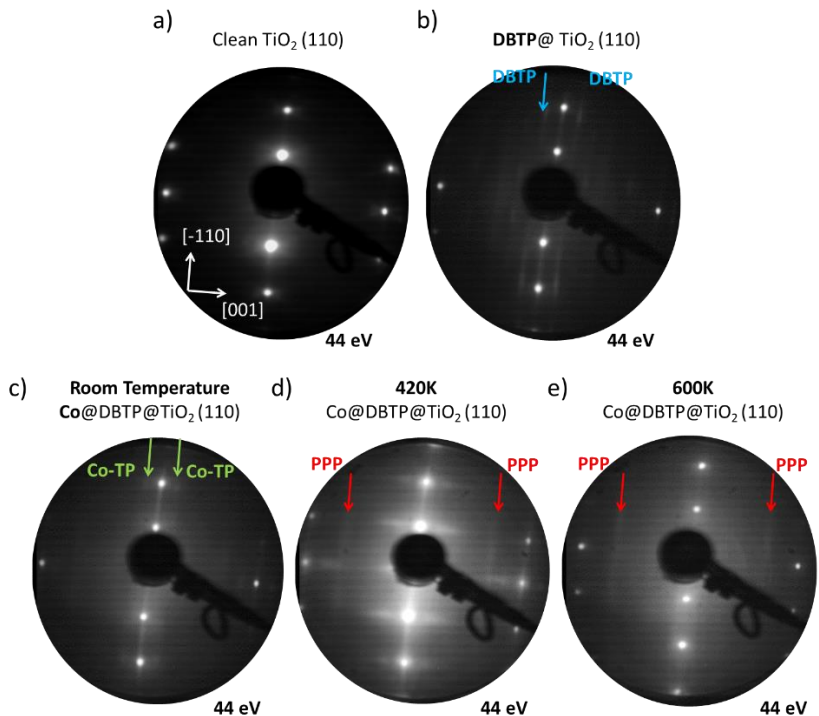

**Figure S 2:** a,b) LEED images taken at 44 eV corresponding to the non reduced $TiO_2$ (110) pristine sample and one monolayer DBTP deposition on the surface. The new diffraction pattern arising from the DBTP molecules is highlighted with blue arrows. c-e) LEED images taken at 44 eV corresponding to the Ullmann coupling reaction of the cobalt sample. At room temperature the diffraction vertical lines corresponding to the organometallic phase (denoted as Co-TP) are observed (panel c). After 420 K annealing step (panel d) new diffraction lines (highlighted with red arrows) show up and arise from the periodicity of the PPP chains. Such PPP chain diffraction patterns persist even after annealing to 600 K (panel e) in agreement with the TD-XPS observations of Figure 2 in the main manuscript. Note that the LEED images are taken at RT after each annealing step.

**STM image of DBTP/Co sample after annealing to 450K**

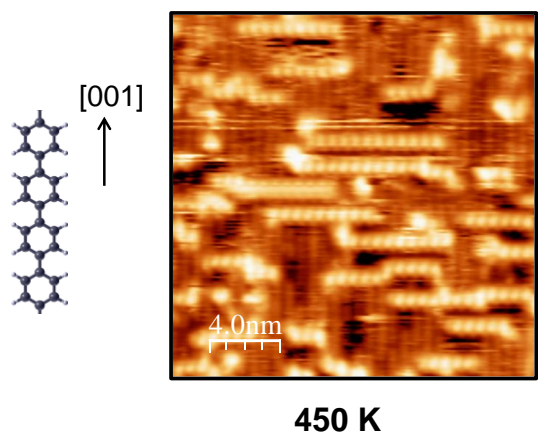

**450 K**

**Figure S 3:** RT-STM image of the cobalt sample obtained after the 450 K annealing step (STM parameters: V= 1,8V, I= 35pA).

The limited surface conductivity of $TiO_2(110)$, restricts the performance of STM as a suitable characterization technique for the current experiment. As a consequence, obtaining accurate images is highly challenging and the obtained results scarce.

The success of using cobalt atoms as catalysts for the Ullmann coupling reaction on $TiO_2(110)$ has been proven by XPS, LEED and ARPES measurements. To gain insight into the formation of the Br-Co complex during the intermediate stages of the reaction, room-temperature scanning tunneling microscopy (RT-STM) was used after annealing the cobalt sample to 450K. In the STM image shown in Figure S3, two distinct structures are resolved perpendicular to each other. On the one hand, with lower intensity, molecular chains are present on the surface along the [001] high-surface anisotropy direction. The polymerization rate and polymer length can not be determined from the images at this stage.

A different, higher intensity structure perpendicular to the [001] direction also appears in the image. This structure has not been reported in previous works in the absence of cobalt atoms, and cannot be attributed to the characteristic $TiO_2(110)$ 2x1 surface reconstruction, as evidenced by the absence of a surface defect state peak in the valence band spectra[2]. Therefore, we correlate the observed structure with the stable Br-Co complex.

**TD-XPS of DBTP and DITP with Co deposition**

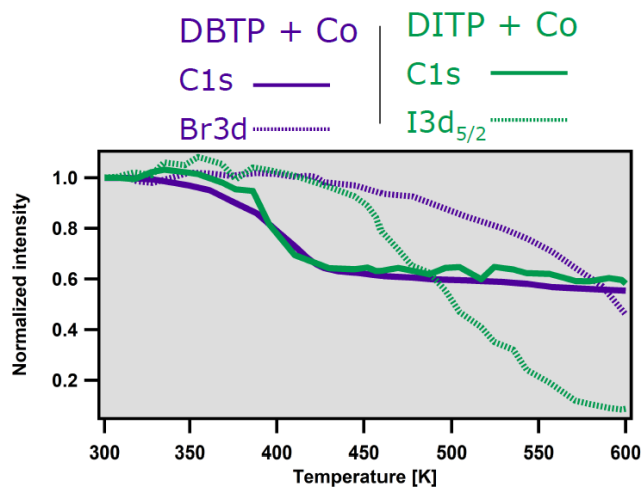

**Figure S 4:** TD-XPS comparison of the normalized intensity areas of the main core level features of DBTP and DITP in the cobalt samples.

Finally, for comparison purposes, we superimpose the normalized intensity spectra measured in Figure 2c and Figure 4c of the main manuscript in Figure S4. The normalized intensity spectra of the C1*s* CL in both molecular precursors, DBTP and DITP, have similar evolution and drop to 60% at 600K, indicating that the Ullmann coupling reaction is practically identical with both precursor molecules after dehalogenation. However, in contrast to the I3*d* signal trend, the Br3*d* signal gradually desorbs and is still present at 600K, likely due to a stronger chemical interaction between cobalt and bromine that prevents bromine from desorbing.